\documentclass[runningheads]{llncs}
\usepackage[colorlinks=true, linkcolor=black, citecolor=blue, urlcolor=blue]{hyperref}
\usepackage{bookmark}
\usepackage[utf8]{inputenc}
\usepackage[T1]{fontenc}
\usepackage{amsmath,amssymb}
\usepackage{booktabs}  
\usepackage{graphicx} 
\usepackage{enumitem}
\usepackage{tikz}
\usepackage{pgfplots}
\usepackage{pgfplotstable}
\usepackage{caption}
\usepackage{xfp}
\usetikzlibrary{positioning}
\pgfplotsset{compat=1.16}
\usepackage{rotating} 
\usepackage[x11names]{xcolor}

\usepackage[x11names,svgnames]{xcolor}
\usetikzlibrary{arrows.meta, shapes.geometric, shapes.misc, positioning, fit, backgrounds}
\usetikzlibrary{calc} 

\definecolor{pRed}    {RGB}{249,200,200} 
\definecolor{pBlue}   {RGB}{192,205,230} 
\definecolor{pYellow} {RGB}{250,237,190} 
\definecolor{pGreen}  {RGB}{199,224,200} 
\definecolor{pPurple} {RGB}{210,200,240} 
\definecolor{pOrange} {RGB}{255,215,170} 

\definecolor{SankeyBlue}{HTML}{1f77b4}
\definecolor{SankeyBlueLight}{RGB}{110,174,236} 
\definecolor{SankeyRed}{HTML}{d62728}
\definecolor{SankeyRedLight}{RGB}{240,140,140}  

\title{An Enactivist Approach to Human-Computer Interaction: Bridging the Gap Between Human
Agency and Affordances}
\titlerunning{An Enactivist Approach to HCI}

\author{Angjelin Hila\inst{1}\orcidID{0009-0005-0481-0443}}

\institute{University of Texas Austin, Austin TX 78712, USA.
\email{ahila@utexas.com}\\}

\begin{document}

\maketitle

\begin{abstract}
Emerging paradigms in XR, AI, and BCI contexts necessitate novel theoretical frameworks for understanding human autonomy and agency in HCI. Drawing from enactivist theories of cognition, we conceptualize human agents as self-organizing, operationally closed systems that actively enact their cognitive domains through dynamic interaction with their environments. To develop measurable variables aligned with this framework, we introduce "feelings of agency" (FoA) as an alternative to the established construct of "sense of agency" (SoA), refining Synofzyk’s multifactorial weighting model and offering a novel conceptual pathway for overcoming gaps in the dominant comparator model. We define FoA as comprising two subconstructs: \textit{affective engagement} and \textit{volitional attention}, which we operationalize through integrated neurodynamic indicators (valence, arousal, cross frequency coupling within the dorsal attention system) and first-person phenomenological reports. We argue that these neurophenomenological indicators provide richer, more actionable insights for digital affordance design, particularly in XR, BCI, Human AI Interaction (HAX), and generative AI environments. Our framework aims to inform and inspire design parameters that significantly enhance human agency in rapidly evolving interactive domains.
\end{abstract}

\keywords{Sense of Agency · Enactivism · Neurophenomenology · Affordances · Interaction Design}

\section{Introduction}
New paradigms are required to further develop the field of human-computer interaction (HCI) in the context of human autonomy and agency, especially in light of emergent HCI paradigms in extended reality (XR), artificial intelligence (AI), and brain-computer interfaces (BCI). We seek to expand understanding of human agents in HCI through grounding within an enactivist theory of cognition. Enactivism proposes that living beings are "autonomous agents that actively generate and maintain themselves by enacting their own cognitive domains" \cite{Thompson2010,Varela1992}. In this paper, we advance a theoretical framework that draws from enactist cognition theory to reconceptualize agents, the environment, and agent-environment interactions within HCI contexts. Our theoretical contributions include proposing alternative means of measuring agency in emerging XR, human-AI interaction (HAX), and BCI modalities and environments. Accordingly, we propose feelings of agency (FoA), operationalized through neurophenomenological indicators, as a more suitable alternative to sense of agency (SoA), which is traditionally measured through the canonical comparator model.

\subsection{Enactivist Cognition and HCI}

Developed by Varela, Rosch, and Thompson, enactivism seeks to bridge the gap between phenomenology and empirical investigation in biology and neuroscience through neurophenomenology \cite{Petitmengin2017}. We argue that this approach requires reconceptualizing the environment as a dynamic search space structured by sociocultural, sensorimotor, and agent-specific affective affordances. Autonomous agents must actively shape this milieu in pursuit of self-realization, rather than merely responding to external inputs. This view challenges the information-processing paradigm, rooted in cybernetic theory, which models agents as input-output systems processing third-person information-bearing states. Instead, we theorize agents as self-organizing systems that instantiate autopoiesis, recursive dynamics that sustain a boundary between agent and environment and facilitate continued viability \cite{MaturanaRumesin1991,Meincke2019,Jaeger2015}. In contrast to third-person information processing, we propose that cognition arises from agent-relative cognitive attractors embedded in historically contingent neural ensembles shaped by patterns of embodied experience \cite{Lewis2005,Buschman2015}

We extrapolate two principles of autopoietic autonomy in the context of human agents: organizational closure and sensorimotor coupling. Organizational closure defines the recursive interdependence between the metabolic network and the agent boundary \cite{Thompson2010,Meincke2019}. Sensorimotor coupling conceptualizes the agent’s relation to the environment as fundamentally normative and affective \cite{vanWieringen2022,Rojas-Libano2019,Thompson2010}. The agent’s perceptual and motile interactions are inextricable from viability constraints entailed by its autopoietic organization. As such, enactivism proposes a view of intentionality that encapsulates object-directedness as well as concomitant coordination of interoceptive and proprioceptive sensations that dynamically enable agents to navigate their search space \cite{Seth2013,Seth2011,Kelso2014}. Embodied perception and action dynamically coevolve toward the realization of agential preferential outcomes as internal to the agent’s sensemaking history. We hypothesize that organizational closure predicts that agents seek feelings of autonomy and agency in HCI environments, whereas sensorimotor coupling predicts that agents enact their own environments by projecting individual relevance criteria in task-positive modes.

In HCI, this framework suggests that agents seek feelings of autonomy and agency when interacting with computational systems. Enactivism generates test\-able hypotheses across multiple timescales of human activity, from attentional dynamics to skill acquisition, and informs the refinement of models such as the circumplex model of affect \cite{Posner2005} and frameworks for skill learning and agent-environment coupling \cite{Tenison2016}. We apply embodied dynamism \cite{Seth2013}, and brain metastability \cite{aguilera2016metastability} to study the coupling of human and computational systems. In what follows, we review related work on agency in HCI, develop an HCI-centric enactivist theoretical framework, reoperationalize the agency variable through neurophenomenological constructs, and derive actionable design parameters for emerging interactive systems. 

\section{Related Work}
Increasing work in HCI recognizes the gap between traditional HCI models and agency. Limerick et al define the gulf of execution as the challenge within HCI to bridge the gap in “form and content” between “the user’s intentions and the systems’ state”\cite{Limerick2014}. Formally, the gulf of execution concerns the transition between psychological states such as intentions to corresponding state changes within the computer with the aim of integrating “initiation requirements, feedback mechanisms, and device capabilities” \cite{Limerick2014}.

Research within implicit SoA in HCI has primarily utilized two measures: intentional binding and sensory attentuation \cite{Limerick2014}. Intentional binding approaches measure the temporal lapse between perceived time of initiation and perceived time of outcome typically using a Libet clock, whereas sensory attenuation paradigms measure the distinction between the predictability of self-generated events and external sensory stimuli using the comparator model \cite{Limerick2014,Blakemore1998,Carruthers2012}, the dominant explanatory model of SoA in cognitive neuroscience. The comparator model views the motor system as a control system with preferred states as inputs and estimated actual states as outputs \cite{Blakemore1998,Synofzik2008}. The brain executes a sequence of motor commands that seek to bridge the gap between internal representations of the desired states and evaluations of actual states \cite{Carruthers2012,Ohata2020}. 

The comparator model leverages the internal model theory and central monitor theory \cite{Blakemore1998,Synofzik2008}. The internal model theory maintains that the central nervous system (CNS) contains knowledge about the sensorimotor system and the external world \cite{Cisek2008,Kawato2007} in order to coordinate actions. The central monitor theory posits a CNS monitoring system that distinguishes self-generated activity from external sensory events by monitoring and evaluating the results of intended actions \cite{Sperry1950,vonHolst1971,Blakemore1998}. The conjunction of these two hypotheses are used to explain SoA. These are typically of two types: forward models and inverse models. Forward models predict the consequences of a given action before it has taken place, whereas inverse models infer the antecedent causes required to achieve an intended effect. 

The ability to monitor and recognize self-generated limb movements is evidenced by the phenomenon of corollary discharge, where an efferent copy of a motor command is broadcast to other regions of the brain allowing them to anticipate self-generated sensory signals and distinguish them from environmental signals \cite{Subramanian2019}. Corollary discharge enables the predictive model to attenuate self-produced sensory feedback to maintain focus on external environmental signals \cite{Subramanian2019}. The general mechanism for coordinating movement involves coordination between the inverse model that computes the required motor movement to transition to a desired state and the forward model that anticipates the consequences of actual movement \cite{Blakemore1998,Synofzik2008}. 

In line with the above hypotheses, three hypothesized comparator mechanisms coordinate voluntary action. Comparator 1, known as the efference copy, compares the difference between desired outcome and actual estimated state \cite{Synofzik2008}. Comparator 2, known as the forward model, compares the difference between the forward model predicted outcome and actual sensory feedback \cite{Synofzik2008}. Comparator 3 utilizes environmental feedback to attenuate accurately predicted sensory inputs \cite{Synofzik2008}. Comparator 1 is hypothesized to account for sense of control, whereas comparator 2 for self-attribution of action. Consequently, the comparator model stakes its definition of sense of agency (SoA) in the congruence between intended outcomes and actual outcomes. Because matching predictions and outcomes do not disrupt agency, the comparator model predicts that the brain discounts the somatosensory consequences of self-generated movements. For this reason, it has been called phenomenologically recessive in the sense that only mismatches between prediction and outcomes produce disruptions in the sense of agency \cite{Buhrmann2017}. 

More recently, aspects of enactivism have gained greater traction within the sense of agency (SoA) literature in cognitive science. Buhrmann and Di Paolo \cite{Buhrmann2017} point out that agency involves the "complex integration of afferent, efferent, and intentional feedback" \cite{Buhrmann2017}. They develop an enactivist theory of sensorimotor agency that integrates dynamical sensorimotor theory, enactive theory of sensorimotor agency, and Piagetian sensorimotor equilibration (SME) \cite{Buhrmann2017}. First, sensorimotor contingency theory (SMCs) posits a non-representational view of action and perception where perceiving is a function of skillful use and practical mastery of the regularities governing worldly interaction. Second, SMCs imply self-producing and maintaining minimal agents that continually individuate themselves against the environment by projecting their intrinsic normativity \cite{Buhrmann2017}. Third, individual sensorimotor schemas differentiate temporally through Piagetian processes of assimilation and accommodation that equilibrate agent normativity, habits, and plans with environmental demands \cite{Buhrmann2017}. While this model assumes the broad congruence of perception and action, it frames the sense of agency in the satisfaction of the ongoing deployment of the agent’s sensorimotor repertoire with respect to its stability and transition dynamics against the demands of the situation. Their model, therefore, does not require precise matching between sensory signals and predicted feedback as posited by the comparator model. 

\section{Gaps in Existing Approaches}
\paragraph{Beyond the Comparator Model.} A major explanatory inadequacy of the comparator model lies with its inability to distinguish between implicit SoA and judgement of agency (JoA). Because the attribution of causal agency in the comparator model occurs at the system or computational level, it cannot explain an agent’s phenomenal experience of agency nor a conceptual judgement of their degree of agency. Analyzing SoA into the feeling of agency (FoA), the phenomenal and non-conceptual awareness of one’s agency, the judgment of agency (JoA), the conceptual attribution of agency to oneself, and the sense of ownership (SoO) as the proprioceptive sense of one’s own body irrespective of whether sensorimotor activity is initiated by the agent, Synofzik demonstrates that efference copy-reafference mismatch is insufficient to either constitute or disrupt agent SoA \cite{Synofzik2008,Synofzik2013}. Patients with parietal lobe lesions who cannot detect mismatches between self-initiated actions and visual consequences report no diminution of JoA and SoA \cite{Synofzik2008}. On the other hand, deafferented patients report a disruption of SoA despite congruence between different copy and sensory feedback \cite{Synofzik2008}. 

A broader limitation of the model is its inability to account for the wide temporal range of intended actions, which span from immediate motor acts like grasping a glass of water to long-term life goals. The comparator model operates at the time scale of sensorimotor movements and perceptual consequences, whereas a global sense of agency monitors a much more protracted time-horizon, monitoring the relationship between micro and macro goals across a spectrum of time spans. Further, by virtue of positing internal models, the comparator model relies on a representational account of mental contents and brain information processing. The degree to which internal representations are imagistic, conceptual or combinations thereof, as well as the degree to which agents have conscious access to them is disputed \cite{Pitt2022,Favela2023}. Contrary to Synofzik, we argue that comparator feedback is necessary but insufficient to explain the feeling of agency (FoA) and judgement of agency (JoA). While we agree with Synofzik that FoA involves a "multifactorial weighting of authorial cues" monitoring the pre-reflective congruence of motor intentions and goals \cite{Synofzik2008}, we explicitly define a new variable termed \textit{feelings-of-agency} (FoA) as task-positive attentional patterns that track endogenous action involving the congruence of the agent's conceptual and phenomenal sense that they are realizing their intrinsic goals. Consequently, we do not explicitly distinguish between conceptual and phenomenal aspects of SoA, but emphasize agent states connected to sensemaking within a continuum of interacting goal-directed time-horizons. 

\paragraph{Beyond the gulf of execution.}
As the gap between user intentions and system states \cite{Limerick2014}, the "gulf of execution" constrains the agency solution-space to bridging the gap between agent intentions and perceived effects in computer states. Solutions include increasing intentional binding by varying the properties of input modalities, increasing feedback reliability, and producing faster command execution to increase sensory attenuation \cite{Limerick2014}. This spectrum of solutions reduces SoA to perceptual-motor control tasks. Because SoA is framed within low-level motor tasks, solutions are geared toward seamless integration of agent actions and system feedback on the one hand, and habituating agent sensorimotor engagement to system affordances, on the other. 

By contrast, our operationalization does not equate SoA with discrete outcomes as construed in the comparator model but with sustained engagement in a context of action in which the agent's actions track a sense of ownership of the process and authorship of provisional outcomes. We expressed this as the process of enacting agent intrinsic norms within the environment. The comparator definition of SoA as agreement between predicted and estimated actual state cannot distinguish between ``doom scrolling" or ``video binging" and modes of engagement that harmonize goal effectance with motivated engagement. As a result, our definition of agency evaluates agent states beyond motor-feedback through \textit{valence}, \textit{arousal} and \textit{volitional attention levels} within a protracted action-context. An operationalization of SoA that tracks \textit{enactive sense of agency} must measure the degree to which agent engagement agrees with their motivational effectance. 

Therefore, according to our analysis, the comparator model is beset by the following limitations: 
\begin{enumerate}
    \item it eschews agent phenomenology 
    \item it omits relevant neurodynamic indicators
    \item it models agency as discrete matching functions rather than dynamical enaction 
    \item it models agents in computational terms as input-output systems
\end{enumerate}

In what follows, we develop a new model of agent-environment relations predicated on the enactivist approach to cognition we have introduced above. Following a reconceptualization of agent-environment relations, we introduce a new variable of agency termed \textit{feelings-of-agency} (FoA). We operationalize FoA by combining phenomenological and neurodynamic indicators, termed neurophenomenology. Our approach provides pathways for improving the comparator model by combining phenomenological analysis of agent experience with neurophysiological indicators. 

\section{An Enactivist Framework for Measuring Feelings of Agency}
Our theoretical approach integrates top-down, first-person description of the phenomenon of personal agency and the bottom-up description of an agent as an autonomous system that realizes its own cognitive domain as a byproduct of its continual self-organization. Therefore, our conceptual framework  merges a) the phenomenological sense of personal agency and b) the bottom-up composition of an autonomous system as realizing the minimal autopoietic organization. The upshot of this integration is that, understood as a property of the whole agent, agency is continuous with the endogenous processes that characterize the autopoietic organization. Autonomy consequently emerges from the endogeny inscribed within the very organization of the agent. This entails that phenomenological experience of agency is continuous with endogenous self-organization. 

\subsection{Redefining Agent-Environment Relations}
In advancing an enactivist approach to HCI, we integrate three foundational theses that describe agent-environment relations: a) \textit{organizational closure}, b) \textit{sensorimotor coupling} and c) \textit{endogenous action}. Jointly, these entail a reconceptualization of the agent as a self-organizing system that endogenously creates meaning, the environment as a continuously enacted relation between endogenous activity and environmental affordances, and the optimal conditions through which autonomous agents interface with one another. 

Unlike the comparator models where the evaluation of agency occurs at critical evaluative junctures between system and environmental states, the enactivist conception of the agent is fluid and the evaluation of outcomes and desired states diffuse with respect to the agent's ongoing neurodynamical skillful and informational coping. In this picture, the agent does not occupy discrete states described by input-output structures, but a continuous action space in which it continually enacts its milieu across dispositional shifts. 

\textbf{Organizational Closure.}
Organizational closure refers to a system of parts connected through closed feedback loops \cite{Thompson2010}. The principle of organizational closure grounds the autonomy of an agent within the self-organizing dynamics of its biological components. This means that agents constitute organizational wholes whose internal dynamics reproduce their identity conditions as a distinct unity across time. Organizational closure distinguishes between merely self-organizing systems, which abound in nature such as convection cells and tornadoes, and autonomous systems, such as "living organisms" that instantiate the autopoietic organization. Drawing from Varella and Maturana \cite{MaturanaRumesin1991}, and Thompson \cite{Thompson2010}, we formalize the necessary and sufficient conditions of a minimal organizational autonomy system as follows: 

\begin{enumerate}[label=\alph*)]
    \item the system produces a semipermeable boundary between itself and the environment
    \item the boundary is reproduced by a network of reactions within the boundary
    \item a) and b) are recursively interdependent
\end{enumerate}

Conditions a) and c) are recursively related as follows: The internal reaction network produces the boundary, while the boundary reproduces the conditions for the reaction network \cite{Thompson2010}. A naturally occurring minimal autopoietic system is a cell. A multicellular organism constitutes a second-order autopoietic system because it is composed of first-order autopoietic systems. All concrete autopoietic systems form subsets of \textit{far-from-thermodynamic equilibrium systems} by continuously reproducing an optimal differential between their internal milieux and external environment \cite{prigogine1946biologie,Thompson2010,zhang2021nonequilibrium}. 

\textbf{Sensorimotor Coupling.}
The conditions for autopoietic organization necessitate that agents exchange matter, energy, and information with their environments \cite{MaturanaRumesin1991,Thompson2010}. This structural feature implies that even a minimal autopoietic system navigates its environment through valanced behaviours of avoidance and approach, namely behaviours that ensure its continued viability in time \cite{dexter2019complex}.  \textit{Sensorimotor coupling}, therefore, refers to the dual ability of an autopoietic system to parse the environment into favorable and unfavorable states through basic sensors and adjust its behaviours to procure nutrients and avoid harm. In more complex agents such as mammals and even more complex ones such as humans that possess language, self-awareness, and mental contents, \textit{sensorimotor coupling} refers to the historical and developmental interactions with the environment constituted as dynamic loops of \textit{attention-perception-action} that result in the \textit{co-determination} of agent and environment.

This view contrasts with the view of cognition as consisting of “information-processing” where a passive agent transduces environmental stimuli into perceptual and higher-order representations \cite{Varela1992,Thompson2010}. In the enactivist framework, information is \textit{constitutionally valanced} for the agent. The notion of constitutionally valanced information countervails the view that cognition can be exhaustively described as the transduction of \textit{third-person}, \textit{objective} information. While physical processes of transduction are indeed components of cognition, they omit informational valence as constitutionally embedded in, and emergent from, the self-organizing dynamics of an autonomous system. Sensorimotor coupling further implies that cognition and action are mutually reinforcing components of agential behavior that subserve its viability against environmental conditions \cite{Rojas-Libano2019,vanWieringen2022}.  Because the agent continually seeks preferential states that subserve its viability, the agent's sensorimotor coupling is consequently inherently normative. Normative coupling regulates sensorimotor activity and, when scaled to human agents, higher-cognitive functions such as planning, goal-orientation, and the production of meaning. Consequently, sensorimotor coupling subsumes “information-processing” within the broader category of “skillful coping”, where skills encompass both sensorimotor activity and higher-order symbol manipulation.  

\textbf{Autonomy \& Endogenous Action}
Our notion of autonomy as grounded in the autopoietic organization can be empirically buttressed by mounting evidence in the brain and physiological sciences for endogenous activity in the nervous system evident even in the simplest organisms such as the hydra \cite{schneider2014brain}. Empirical analysis of simple nervous systems indicates that, unlike the passive input-output activation model, the brain mediates even basic stimulus-response scenarios with modulatory circuits \cite{schneider2014brain,brembs2021brain,barabasi2024functional}. This evidence advances a view of the brain as a dynamically active organ \cite{brembs2021brain}, which reframes cognition and action as spontaneously arising from within the organism rather than as merely reacting to environmental signals. As a dynamically active system, the brain leverages learning and plasticity to expand, differentiate, and complexity its native circuitry \cite{barabasi2024functional}. We posit that the brain leverages metastability in order to settle into and transition from attention basins, where metastability refers to a state of a system that is locally stable under small disturbances but can transition to a more stable state if sufficiently perturbed \cite{aguilera2016metastability}. 

With respect to goal-oriented action and task-positive engagement modes, we argue that attentional basins evolve idiosyncratically for human agents relative to their developmental history. Lewis's \cite{Lewis2005} dynamic system's approach to emotional organization provides supporting evidence for this account by demonstrating that cognition and emotion co-evolve through bidirectional emotion-appraisal patterns that generate organizational states of self-amplification of self-stabilization. We argue that these organizational dynamics play a crucial role within HCI contexts, shaping the agent's ongoing interaction with digital affordances. Affordances can amplify the agent's attentional self-organization by propelling them in a state of deep engagement and creativity or they can dissolve attentional self-organization through negative-feedback loops and increase of uncertainty. We hypothesize that when an agent's endogenous motivation is met with negative valence and deferred rewards, it can result in a diminution of agency and long-term withdrawal from particular sets of digital affordances. Ongoing interaction with digital affordances, therefore, shapes an agent's pattern of self-regulation and tendencies for endogenous activity. 

\textbf{Agent-Environment Reconstitution.}
Our refinement of the enactivist framework entails a reconceptualization of the environment. We reconceptualize the environment as agent-relative and dynamic rather than fixed and objective, and re-operationalize it as a field of action constituted by sensorimotor and affective coupling. The same external conditions can constitute distinct environments for different agents, shaped by feedback loops that reinforce individualized patterns of engagement.  Each agent enacts their own environment as a function of its own self-organized emotion-appraisal patterns that propel them toward the future as a function of self-maintenance. In this sense, the organism's environment constitutes its \textit{umwelt} or environmental niche, defined as the process through which organisms actively modify their environments, creating co-evolutionary agent-environment feedback loops \cite{odlingsmee1996niche}. 

We argue that HCI contexts increasingly form the predominant means through which agents pursue personal and professional effectance. Consequently, computational extensions extend the agent's sensorimotor coupling through input modalities but also reconstitute the environment  by overlaying digital spaces that require new skill acquisition and internal modeling. The habitual feedback loops that emerge from ongoing agent interaction with computational extensions and GUI interfaces, continually reproduce the environment as those salient patterns of perception-attention-action.  HCI contexts, consequently, present ways that agents can augment their agency through greater effectance but also ways that agents can lose agency through action reward-patterns that derail them from effortful, intentional action. Examples of the latter include video binging, and social media and video game addiction. In these latter cases, environmental affordances co-opt endogenous user action into short-term reward patterns resulting in a loss of effectance and feelings of agency. Figure 1 illustrates the dynamic loops of agent sensorimtor action and environmental feedback as an ongoing mutual reconstitution of agent and environment. 

\begin{figure}[h]
    \centering
    \includegraphics[width=0.8\textwidth]{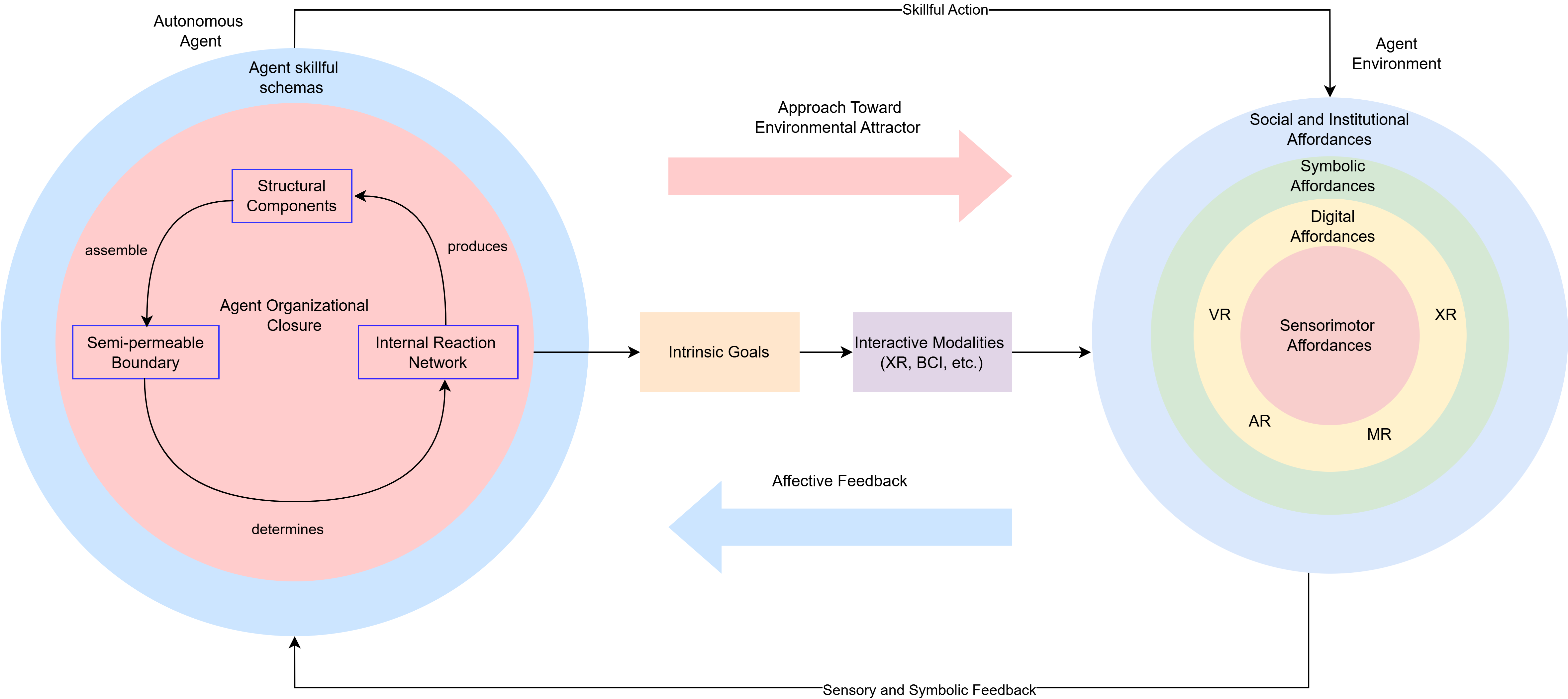}
    \caption{Agent-Environment Reconstitution}
    \label{fig:Agent-Environment Model}
\end{figure}

\subsection{From Sense of Agency to Feelings of Agency}
To operationalize human agency in HCI, we develop a construct of feelings of agency that departs from Synofzik et al.’s decomposition sense of agency (SoA) into judgment of agency (JoA) and and feeling of agency (FoA). While Synofzik delineates these into separate dimensions, emphasizing self-causation, we instead propose an integrated account. Our construct blends pre-reflective affective experience with higher-order cognitive regulation and intentional direction. Rather than distinguishing “feelings” from “judgments,” we conceptualize feelings of agency as a dynamic composite: a phenomenologically rich, ongoing sense of authorship and engagement shaped through both embodied interaction and reflective modulation. 

In line with our enactivist account, we define FoA as a composite construct encompassing two key dimensions: (1) Affective Engagement (AE), the degree of positive arousal, emotional involvement, and perceived meaningfulness of action, and (2) Volitional Attention (VA), the sustained and intentional orientation toward action outcomes. Phenomenologically, affective engagement tracks the meaningfulness of the task for the agent, whereas volitional attention tracks the degree to which the agent feels authorship for the task. These dimensions reflect the enactive premise that agency emerges from the coordination of perception, attention, and affect in relation to endogenous goals. Below, we provide the requisite evidence that supports these constructs. 

\textbf{Attentional Self-Organization.} Neurodynamics research offers growing empirical support for these constructs. Task engagement has been linked to activity in the task positive network (TPN), particularly in the dorsal attention system (DAS), responsible for volitional attention \cite{vossel2014dorsal,fox2006spontaneous}. DAS activity corresponds with higher frequency bands such as beta (13–30 Hz) and gamma (30–100 Hz)\cite{beste2023towards}, while disengagement correlates with default mode network (DMN) activity in lower frequency bands such as theta (0.4–4 Hz) and alpha (8–13 Hz)\cite{raichle2015,klimesch2018}. Evidence further indicates that DAS coordinates between bilateral dorsal attention and right-lateralized ventral attention in order to flexibly transition between attention states \cite{Lewis2005,fox2006spontaneous}. In particular, the orbitofrontal cortex (OFC) and anterior cingulate cortex (ACC) coordinate with the amygdala and ventral striatum to mediate both pre-attentive and attentive appraisal \cite{Fox2006}. The OFC encodes and sustains attention to context-specific, motivational contingencies and, in coordination with the amygdala, monitors changes in the hedonic valence of anticipated events \cite{Lewis2005}. While both dorsal and ventral ACC regions contribute to emotion and attention, the dorsal ACC is specifically involved in explicit evaluations of voluntary choice and directed learning \cite{Lewis2005,Chen2024}. Tognoli and Kelso's theory of brain metastability, where partial synchronization and frequency locking patterns balance global integration with local segregation, provides a general framework for measuring this activity \cite{tognoli2014enlarging}. 

\textbf{Attention \& Emotion.} We draw from Lewis's emotion self-organization theory~\cite{Lewis2005,lewis2007self} of emotional dynamics to ground the role of affect in attentional and motivational processes. Lewis proposes that emotional regulation emerges from interacting time-scales of real-time emotional development, the mesodevelopment of moods across hours and days, and the macrodevelopment of personality across years \cite{lewis2000emotional}. Lewis further posits that cognitive appraisal and emotion reciprocally produce neurodynamical patterns that regulate directed attention in real-time \cite{Lewis2005}. 

\textbf{Valence \& Arousal.} Given the intimate connection between attention and emotion, we propose utilizing valence and arousal as variables that measure affective engagement. Valence refers to the positive or negative direction of an emotional experience, while arousal describes the intensity or activation level of that experience ~\cite{Storbeck2008,Basu2015,Yik2022}. Jointly these regulate approach and withdrawal tendencies and dynamically shape how attention is allocated and sustained during task engagement. Emotional arousal and motivational valence, regulated through systems the anterior cingulate cortext (ACC) and frontal alpha asymmetry (FAA) \cite{Sabu2022}, shape attentional states and decision-making, consistent with Lewis's dynamic systems model of emotion-appraisal loops \cite{Lewis2005}.

\textbf{Neurophenomenology.} Together, these insights support a model in which FoA can be operationalized using neurodynamic indicators such as cross frequency coupling (CFC) in the anterior cingulate cortex (ACC) and orbitofrontal cortex (OFC), frontal-alpha asymmetry and alpha-beta ratios, in combination with phenomenological self-reports that reflect engagement, authorship, and emotional resonance. We thus propose a neurophenomenological framework in which phenomenological data~\cite{Berkovich-Ohana2020} provide the concrete means for linking subjective experience with the underlying neurodynamics of agency.

\subsection{Operationalizing Feelings of Agency} Building on the preceding theoretical framework, we now provide a structured operationalization of feelings of agency (FoA) suitable for empirical investigation in task-positive HCI contexts. We operationalize each construct through neurodynamic and phenomenological indicators. We define affective engagement as \textit{positively valenced arousal} and Sense of Authorship as \textit{sustained voluntary attention} during tasks. 

\textbf{Affective Engagement.} We define affective engagement as a composite of valence and arousal indicators, measuring the intensity and hedonic tone of experience. We measure arousal through alpha-beta ratio \(\alpha/\beta\) and valence through frontal-alpha asymmetry (FAA) \cite{beste2023towards}. Phenomenologically, we hypothesize that positive arousal will correspond to phenomenological reports of emotional involvement, deep engagement, and novelty. 

\textbf{Volitional Attention.} We propose measuring sustained voluntary attention through activity in the anterior cingular cortex (ACC) and orbitofrontal cortex (OFC) \cite{kondo2004cooperation,rossi2009prefrontal,lovstad2012anterior}. Specifically, activity in these areas can be measured through a type of cross-frequency coupling (CFC) called theta-gamma \textit{phase-amplitude coupling} (PAC) \cite{daume2024control}. The dorsal regions of the ACC are correlated with explicit evaluation of voluntary choice and directed attention \cite{Lewis2005}. The OFC, conversely, monitors ``hedonic valence" of attentional contingencies \cite{Lewis2005,fox2006spontaneous,vossel2014dorsal}. In conjunction with other brain areas like the amygdala, the ACC and OFC together regulate directed attention and moment-to-moment relevance appraisal \cite{fox2006spontaneous,vossel2014dorsal}. We hypothesize PAC in the ACC and OFC will correspond to phenomenological reports of ownership, authorship, and intended actions. 

Together, these neurodynamic measures indicate "tendencies to approach" and "tendencies to withdraw" in task-positive contexts. Their hypothesized correspondence with phenomenological reports can help test whether brain activity in these areas correlates to first-person agent experience. Figure 2 displays the operationalization of FoA into the constructs of affective engagement and volitional attention and their corresponding neurodynamic and phenomenological indicators. Table 1 displays an experimental design that measures FoA in a virtual reality (VR) context with task openness and agent motility as independent variables. 

\begin{table}[h]
 \resizebox{\textwidth}{!}{%
\centering
\begin{tabular}{|p{3.3cm}|p{5.5cm}|p{5.5cm}|}
\hline
\textbf{Construct or Dimension} & \textbf{High-Agency Task} & \textbf{Low-Agency Task} \\
\hline
\textbf{Agency} & High & Low \\
\hline
\textbf{Task Type} & Conceptual generation with interactive feedback & Recognition-based response in constrained format \\
\hline
\textbf{Motility} & High (semantic, verbal, perspective-taking) & Low (visual/gaze-based, minimal movement) \\
\hline
\textbf{Openness} & Flexible, Endogenous initiation & Inflexible, Externally guided \\
\hline
\textbf{Attention Profile} & Broad: exploratory, semantic, multi-focus & Narrow: reactive, recognition-driven \\
\hline
\textbf{Cross-Frequency Coupling (CFC)} & Elevated (ACC/OFC coordination of volitional control) & Reduced (minimal intentional engagement) \\
\hline
\textbf{Alpha/Beta Ratio} & Low alpha, high beta (active exploration) & Elevated alpha, sustained beta (focused compliance) \\
\hline
\textbf{Frontal Alpha Asymmetry (FAA)} & Left-dominant (approach-oriented) & Neutral or right-dominant (compliance/avoidance) \\
\hline
\textbf{Phenomenological Indicators} & Strong authorship, ownership, affective engagement & Reduced authorship, novelty, and emotional salience \\
\hline
\end{tabular}
}
\caption{Theoretical contrast between high-agency and low-agency tasks across attentional, affective, neurodynamic, and phenomenological dimensions.}
\end{table}

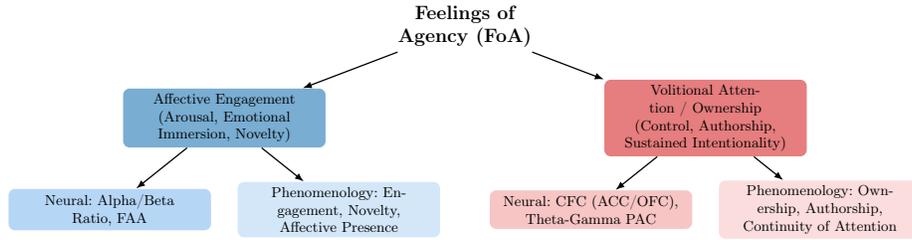
\begin{figure}[h]
\centering
\resizebox{\textwidth}{!}{
\begin{tikzpicture}[
  every node/.style={font=\small, text width=4.2cm, align=center},
  level distance=2cm,
  sibling distance=10.5cm,
  edge from parent/.style={draw, -{Latex[length=2mm]}, thick},
  grow=down,
  level 2/.style={sibling distance=5cm}
]
\node[font=\bfseries\large] {Feelings of Agency (FoA)}
  child { node[fill=SankeyBlue!60, rounded corners] {Affective Engagement \\ \footnotesize (Arousal, Emotional Immersion, Novelty)}
    child { node[fill=SankeyBlueLight!50, rounded corners] {Neural: Alpha/Beta Ratio, FAA} }
    child { node[fill=SankeyBlueLight!30, rounded corners] {Phenomenology: Engagement, Novelty, Affective Presence} }
  }
  child { node[fill=SankeyRed!60, rounded corners] {Volitional Attention / Ownership \\ \footnotesize (Control, Authorship, Sustained Intentionality)}
    child { node[fill=SankeyRedLight!50, rounded corners] {Neural: CFC (ACC/OFC), Theta-Gamma PAC} }
    child { node[fill=SankeyRedLight!30, rounded corners] {Phenomenology: Ownership, Authorship, Continuity of Attention} }
  };
\end{tikzpicture}}
\caption{Hierarchical decomposition of Feelings of Agency (FoA) into Affective Engagement and Volitional Attention/Ownership with corresponding neural and subjective indicators.}
\end{figure}

\section{Design Implications and Applications}
\subsection{Design Parameters}
In order to develop enactivist-derived design parameters, we employed a deductive concept-driven design mapping \cite{Hook2009,dalsgaard2014bridging}, starting from enactivist principles, moving through activity-centered constructs, identifying measurable affective-neurodynamic indicators, and operationalizing these as actionable design parameters targeted to emerging HCI domains (XR, BCI, AI-HCI). We derived three broad design parameters: (a) engagement-disengagement modulation, (b) Affordance gradation and (c) Expressive Interfaces. 

\begin{sidewaysfigure}
    \centering
    \includegraphics[width=1.1\textwidth]{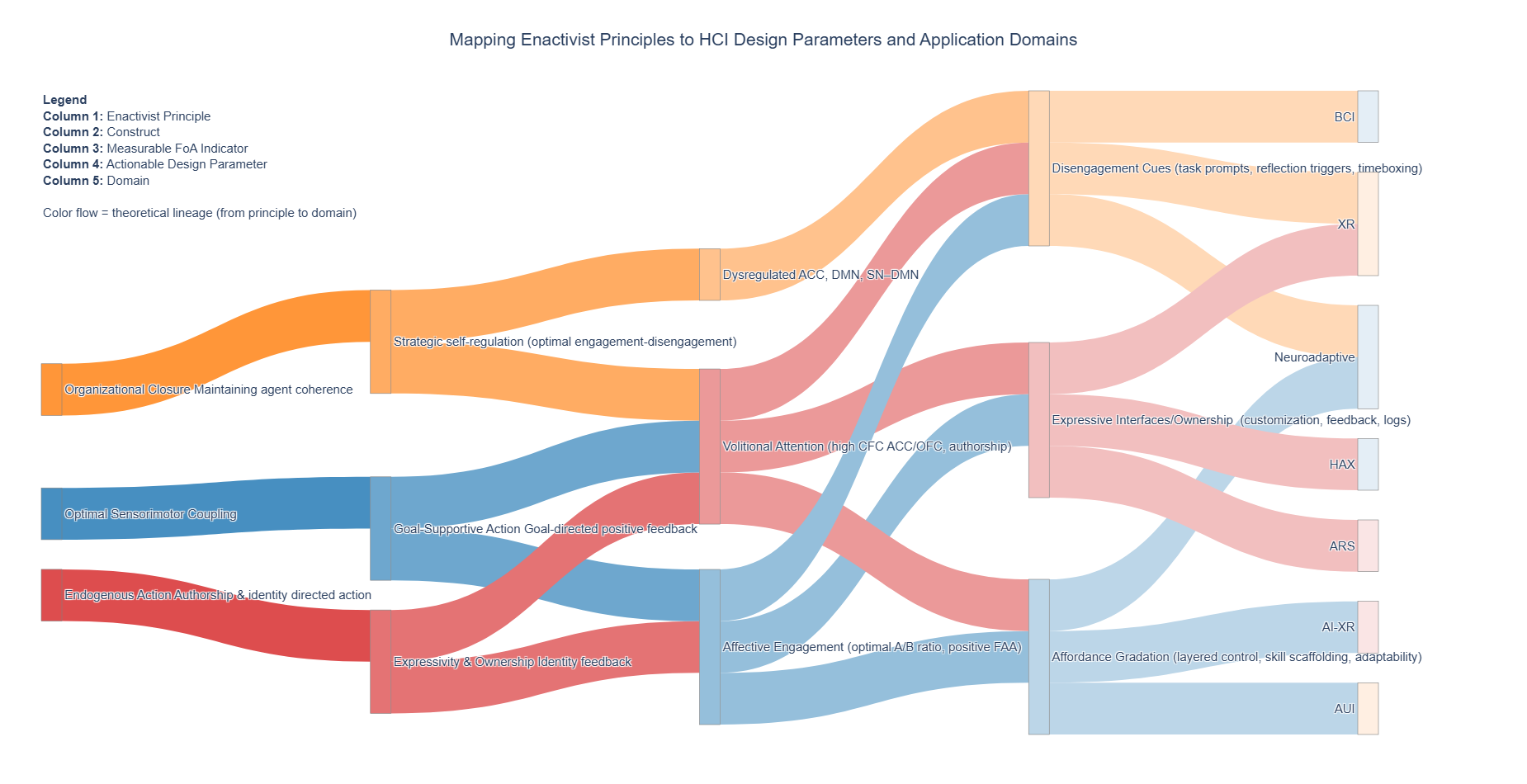}
    \caption{...}
    \label{fig:enactivist-sankey}
\end{sidewaysfigure}

\textbf{Engagement-Disengagement Modulation.}
Disengagement cues are a direct application of the enactivist principle of organizational closure, framing agent autonomy as an ongoing process of self-maintenance that requires dynamic modulation between engagement and disengagement. Human agents operate across multiple regulatory domains—bodily, cognitive, and social—that necessitate periodic withdrawal from task-focused (central executive network, CEN) states to internally oriented (default mode network, DMN) or salience network (SN) states for restoration and reflection. In HCI contexts, excessive unbroken engagement typified by addictive social media use, binge gaming, or unending immersive XR experiences can dysregulate ACC responses, produce DMN hyperconnectivity, and disrupt the balance of SN–DMN interaction. These disruptions are linked to attentional fatigue, reduced volitional control, and ultimately diminished autonomy.

From a design standpoint, actionable disengagement cues include task completion prompts, reflection triggers, and adaptive thresholds that gently suggest disengagement after extended interaction or rising error rates. For example, an AI-driven writing assistant might detect sustained effort and, upon noticing fatigue through patterns such as repetitive errors or long pauses, recommend a mental reset. In XR environments, the system could fade the scene or introduce reminders to take a break after periods of high arousal or continuous action, while in BCI-based interfaces, neuroadaptive algorithms could identify markers of cognitive overload and prompt disengagement. These cues are calibrated to the agent’s internal states, designed to conserve autonomy, replenish attentional resources, and prevent the transition from active, goal-driven engagement to passive, habitual, or exploitative loops. By linking real-time neurodynamic indicators such as dysregulated ACC or SN–DMN balance and phenomenological states like fatigue or loss of focus to adaptive disengagement cues, HCI systems in XR, HAX, AI-driven, and BCI domains can foster a sustainable, agent-centered mode of interaction that aligns with biological and cognitive imperatives for self-maintenance.

\textbf{Affordance Gradation.}
Affordance gradation is grounded in the enactivist principle of sensorimotor coupling and supports intentional action by progressively aligning system complexity with user skill and goals. While sensorimotor coupling is the default state for the agent, its quality can either support or undermine agency. Negative feedback loops, such as rigid interfaces or poorly calibrated feedback, can generate frustration, fatigue, or disengagement, evidenced neurodynamically in blunted affective engagement (such as reduced FAA asymmetry or suboptimal alpha/beta ratios) and phenomenologically in boredom or confusion. By contrast, positive sensorimotor coupling supports affective engagement and volitional attention, as indicated by optimal frontal alpha asymmetry and high engagement self-reports. Affordance gradation operationalizes these principles by scaffolding control and feedback, offering layered interaction modes that users can unlock or customize as their skill develops. For example, in XR-based physical rehabilitation, systems may begin with simple, constrained movements, gradually enabling more nuanced manipulation and user-driven goal setting as competence increases. In adaptive learning environments, scaffolded skill acquisition appears as stepwise tutorials or unlockable features, like a VR sculpting app that introduces basic forms before advanced tools. Adjustable interaction rhythms further optimize coupling, allowing users to control the tempo of engagement; an AI writing assistant, for example, might allow users to pause after each suggestion or shift to rapid brainstorming as desired. By allowing agents to modulate both engagement tempo and complexity, affordance gradation ensures system feedback is maximally aligned with user intentions, supporting autonomy, learning, and affective engagement across XR, BCI, and adaptive AI domains.

\textbf{Expressive and Authorial Interfaces.}
Expressive or authorial interfaces are rooted in the enactivist principle of endogenous action, emphasizing self-generated and identity-relevant activity over passive consumption. Endogenous action is closely linked to volitional attention, measurable neurodynamically by high cross-frequency coupling in regions like the ACC and OFC, which index sustained intentional engagement and context-sensitive authorship. Phenomenologically, this manifests as reports of creative flow, ownership, and satisfaction with outcomes. In HCI, expressive interfaces extend rather than replace agent initiative. For instance, in generative AI art platforms, the human agent remains the primary creator and initiator of action, dynamically modulating how and to what extent generative AI contributes. Users input prompts, iteratively adjust parameters such as automatic completion and color palette suggestion, and steer the generative process, with the system logging choices to reinforce agency and provide an authorship trace. In XR domains, expressive interfaces similarly empower users to build immersive worlds through gestures or movement, as in sandbox games where users assemble, manipulate, and program interactive elements. BCI-based systems further extend authorial control by mapping neurodynamic intention or affect to creative parameters, such as a music composition tool modulating harmony in response to real-time alpha asymmetry or attention state. These systems thus position users as active authors, enabling flexible interaction that aligns dynamically with evolving intentions and supports ongoing learning and self-expression.

\subsection{Future Applications}
\textbf{Extended Reality (XR).}
Our conceptualization of digital affordances as an extension of the agent-environment dynamic coupling necessitates the incorporation of emerging modalities that further transform this relationship. The term extended reality (XR) refers to ``computer-generated immersive environments that provide a spectrum of experiences, including: augmented reality (AR), mixed reality (MR), virtual reality (VR), and XR blended" \cite{stanney2021xr}. Stanney et al \cite{stanney2021xr} argue that XR is "becoming a common-place tool for training and educating, collaborating and partnering, networking and entertaining, and other applications which have yet to be envisioned and realized". AR refers to the overlaying of virtual content onto the real world; MR additionally enables the virtual content to interact with the real world; VR immerses users in a fully virtualized world; and finally XR blends all modalities \cite{stanney2021xr}. In XR contexts, the enactivist framework enables a rethinking of design strategies beyond immersion toward embodied agency. Drawing from the dimensions in Table 1, we propose concrete XR implementations that enhance volitional attention, affective engagement, norm-realizing autonomy, authorship, and sensorimotor coupling. For example, XR systems can support volitional attention through scalable interface complexity, while affective engagement can be modulated through real-time bioadaptive environments. These dimensions converge in designing XR systems that are not merely reactive but co-regulative with the user, enabling sustained and meaningful forms of agency within extended environments.

\textbf{Brain-Computer Interfaces (BCI).}
Originally developed for assistive technologies, Brain-Computer Interfaces (BCIs) enable direct neural control over digital systems, replacing or augmenting motor output in individuals with neuromuscular disorders \cite{he2020bci}. Recent advances expand BCI applications to the general population, enhancing attention, learning, media interaction, and expressive communication \cite{he2020bci}. When combined with XR, BCI offers a powerful substrate for enactive agency by coupling neurodynamic feedback with immersive environments. For example, neuroadaptive XR can support volitional attention through real-time modulation of interface complexity, or affective engagement through bio-signal-driven ambiance and feedback. In accessibility contexts, BCI-XR systems can simulate embodied interaction, enabling users to enact intentions and norms even in the absence of full motor function. For broader populations, BCI-XR may facilitate more intuitive, affectively transparent, and self-generated interaction through the integration of implicit emotional signals and explicit intentional commands. This convergence invites new paradigms of human-computer coupling where embodied cognition and agency are co-constructed across neural and virtual domains.

\textbf{AI-HCI Integration.}
The emergence of large language model (LLM) chatbots has given rise to a new domain within HCI known as human-AI interaction (HAX) \cite{jarrahi2022hybrid}. Our enactivist operationalization of Feelings of Agency (FoA) provides a novel framework for empirically evaluating which forms of HAX interaction enhance agency and which risk displacing it. While LLM applications are rapidly proliferating, little empirical research has examined how different interaction types affect users’ sense of competence, volitional engagement, or cognitive offloading. Consider two illustrative scenarios: (a) a user delegates essay writing entirely to a chatbot and makes minor edits, and (b) a user prompts the chatbot for ideas but writes the essay independently. These interactions exist on a spectrum of agency distribution. Current models such as the comparator model, which focus narrowly on predictive error minimization, cannot distinguish between these interaction types. In contrast, our enactivist FoA model identifies agency with norm realization, wherein authorship emerges through self-generated, goal-directed interaction that aligns with intrinsic preferences and embodied sense-making. Thus, the phenomenological character and neurodynamic signature of the interaction, not merely its functional success, determine whether agency is preserved, augmented or reduced in human-LLM interaction.

\textbf{AI-XR Integration.}
AI-XR refers to the integration of artificial intelligence into the generation, personalization, and dynamic adaptation of XR environments. With advances in generative models, particularly in scene synthesis, affective computing, and procedural content generation, AI is poised to become a primary agent in the co-construction of virtual milieus. Within an enactivist framework, this opens possibilities for norm-realizing autonomy, where users actively shape their sensory and social environments to align with intrinsic values and creative intentions. AI-driven personalization may augment agency by enabling users to enact and inhabit virtual worlds that reflect their preferences, goals, and imaginative capacities. However, the same systems may also lead to experiential fatigue, over-personalization, and forms of cognitive closure that isolate users from shared, intersubjective realities. Enactivist HCI must therefore critically evaluate whether AI-XR systems scaffold meaningful sense-making or trap users in self-referential, disembodied feedback loops. Design frameworks should emphasize co-adaptive affordance modulation, collaborative norm expression, and hybrid human-AI authorship to preserve the richness and openness of user agency.

\section{Conclusion}
In this paper we have laid the groundwork for an enactivist cognition approach to HCI. In doing so, we have operationalized agents as dynamical, self-organizing, autonomous systems that continually reproduce their viability conditions. The dominant models that measure sense of agency (SoA) in HCI contexts are oriented toward solving the gulf of execution, namely the local effectance gap between agent actions and computational feedback. This problem-framing orients usability and interaction design research in current and emerging HCI paradigms spanning GUI and direct-manipulation interfaces, HAX, AUI, XR, and BCI, toward seamless integration of human agent and system.  We have argued that this manner of framing the problem overlooks the broader definition of human agency, which involves utilizing HCI affordances to enact endogenous goals. Current models of HCI sensorimotor-coupling can lead to forms of interaction that, while achieving local integration, diminish user agency by embroiling them in self-reinforcing compulsive interactions that prioritize short-term rewards. Consequently,  evolving HCI environments present unique threats to human agency by co-opting it into behaviors that subserve short-term rewards over long-term goal realization. Because the comparator model is ill-equipped to distinguish between these two types of agency, we believe that combining neurodynamic and phenomenological measures can bridge this explanatory gap. Leveraging neurophenomenology and enactivist insights, we have reoperationalized SoA into the variable feelings of agency (FoA), measured through the constructs of \textit{affective engagement} and \textit{volitional attention}, and mapped these onto design parameters that aim to conserve and augment agent authorship within XR, BCI and Gen AI environments. This operationalization opens up possibilities for empirical studies into burgeoning XR, BCI, and AI-HCI integration. 

\bibliographystyle{splncs04}
\bibliography{bibliography}

\end{document}